\documentstyle[twoside,fleqn,espcrc2]{article}

\newcommand{\be}{\begin{equation}}
\newcommand{\ee}{\end{equation}}
\newcommand{\bea}{\begin{eqnarray}}
\newcommand{\eea}{\end{eqnarray}}
\newcommand{\bfg}{\begin{figure}[htb]}
\newcommand{\efg}{\end{figure}}
\newcommand{\lb}{\label}
\newcommand{\rf}{\ref}

\newcommand{\AmS}{{\protect\the\textfont2
  A\kern-.1667em\lower.5ex\hbox{M}\kern-.125emS}}
\title{Pionium lifetime and $\pi\pi$ scattering lengths in 
       generalized chiral perturbation theory}  

\author{H. Sazdjian\address{Groupe de Physique Th\'eorique, 
    Institut de Physique Nucl\'eaire,\\
    Universit\'e Paris XI, F-91406 Orsay Cedex, France\\
    E-mail: sazdjian@ipno.in2p3.fr}}

\begin{document}

\begin{abstract}
The relationship between the pionium lifetime and the $\pi\pi$ scattering 
lengths is established, including the sizable electromagnetic corrections.
The bound state formalism that is used is that of constraint theory which
provides a covariant three-dimensional reduction of the Bethe--Salpeter 
equation. The framework of generalized chiral perturbation theory allows
then an analysis of the lifetime value as a function of the $\pi\pi$ 
scattering lengths, the latter being dependent on the quark condensate value.
\end{abstract}

\maketitle

The possible measurement of the pionium ($\pi^+\pi^-$ atom) lifetime with
a 10\% precision in the DIRAC experiment at CERN \cite{dir} is expected to
allow a determination of the combination $(a_0^0-a_0^2)$ of the $\pi\pi$
scattering lengths with 5\% accuracy; here, $a_0^I$ is the strong 
interaction (dimensionless) $S$-wave scattering length in the isospin
$I$ channel.
The strong interaction scattering lengths $a_0^0$ and $a_0^2$
have been evaluated in the literature in the framework of chiral
perturbation theory ($\chi PT$) to two-loop order of the chiral effective
lagrangian \cite{gl,bcegs,kmsf}. Therefore, the pionium lifetime measurement
provides a high precision experimental test of chiral perturbation theory
predictions.
\par
The nonrelativistic formula of the pionium lifetime in lowest order of
electromagnetic interactions was first evaluated by Deser \textit{et al.}
\cite{dgbt}. It reads:
\be \lb{e1}
\frac{1}{\tau_0} = \Gamma_0 = \frac{16\pi}{9}\sqrt
{\frac{2\Delta m_{\pi}}
{m_{\pi^+}}} \frac{(a_0^0-a_0^2)^2}{m_{\pi^+}^2} |\psi_{+-}(0)|^2,
\ee
where $\Delta m_{\pi}=m_{\pi^+}-m_{\pi^0}$ and $\psi_{+-}(0)$ is the 
wave function of the pionium at the origin (in $x$-space).
\par
A precise comparison of the theoretical values of the strong interaction
scattering lengths with experimental data necessitates, however, an
evaluation of the corrections of order $O(\alpha)$ ($\alpha$ being the
fine structure constant) to the above formula. Such an evaluation was
recently done by several authors. In the frameworks of quantum field
theory and $\chi PT$, three different methods of evaluation have led to 
the same estimate, of the order of 6\%, of these corrections
\cite{js,s,illr,gglr}. The first method uses a three-dimensionally reduced
form of the Bethe--Salpeter equation (constraint theory approach) and deals 
with an off-mass shell formalism \cite{js,s}. The second method uses the
Bethe--Salpeter equation with the Coulomb gauge \cite{illr}. The third one
uses the approach of nonrelativistic effective theory \cite{gglr}.
\par
The pionium lifetime, with the sizable $O(\alpha)$ corrections included in,
can be represented as:
\bea \lb{e2}
\frac{1}{\tau}=\Gamma&=&\frac{1}{64\pi m_{\pi^+}^2}
\big({\cal R}e\widetilde{{\cal M}}_{00,+-}\big)^2 (1+\gamma)
\nonumber\\
& &\times |\psi_{+-}(0)|^2 \sqrt{\frac
{2\Delta m_{\pi}}{m_{\pi^+}}(1-\frac{\Delta m_{\pi}}{2m_{\pi^+}})}
\nonumber \\
&\equiv& \Gamma_0\sqrt{\Big(1-\frac{\Delta m_{\pi}}{2m_{\pi^+}}\Big)}
\Big(1+\frac{\Delta \Gamma}{\Gamma_0}\Big),
\eea
where ${\cal R}e\widetilde{{\cal M}}_{00,+-}$ is the real part of
the on-mass shell scattering amplitude of the process $\pi^+\pi^-\rightarrow
\pi^0\pi^0$, calculated at threshold, in the presence of electromagnetic
interactions and from which singularities of the infra-red photons have
been appropriately subtracted \cite{ku}; the factor $\gamma$ represents 
contributions at second-order of perturbation theory with respect to the 
nonrelativistic zeroth-order Coulomb hamiltonian of the bound state
formalism. The explicit expressions of
${\cal R}e\widetilde{{\cal M}}_{00,+-}$ and of $\gamma$ may differ
from one approach to the other, but their total contribution should be the
same.
\par
Our aim is to extend the previous analysis to the
case of generalized $\chi PT$ \cite{fss,kmsf}. The latter is based on
the observation that the fundamental order parameter of spontaneous chiral
symmetry breaking is $F_{\pi}$, the decay coupling constant of the pion,
which is related to the two-point function of left- and right-handed
currents in the chiral limit. The other order parameters, such as the quark
condensate in the chiral limit, $<0|\overline qq|0>_0$, have values 
depending on the details of the mechanism of chiral
symmetry breaking and require independent experimental tests. Standard
$\chi PT$ is based on the assumption that the value of the 
Gell-Mann--Oakes--Renner (GOR) parameter \cite{gmorgw}, defined as
\be \lb{e3}
x_{GOR}=-\frac{2\hat m<0|\overline qq|0>_0}{F_{\pi}^2m_{\pi}^2},
\ee
where $2\hat m=m_u+m_d$ and $m_{\pi}$ is the pion physical mass, is close 
to one. Stated differently, the quark condensate parameter 
\be \lb{e4}
B_0\equiv -\frac{<0|\overline qq|0>_0}{F^2},
\ee
where $F$ is $F_{\pi}$ in the chiral $SU(2)\times SU(2)$ limit, 
is of the order of the hadronic mass scale $\Lambda_H\sim 1$ GeV. This
assumption fixes the way standard $\chi PT$ is expanded: the quark
condensate parameter $B_0$ is assigned dimension zero in the infra-red 
external momenta of the Goldstone bosons, while the quark masses are 
assigned dimension two \cite{gl}.
\par
Generalized $\chi PT$ relaxes the previous assumption and treats the
order of magnitude of the quark condensate parameter $B_0$ as an
\textit{ a priori} unknown quantity (awaiting a precise experimental
information about it) leaving to it the possibility of reaching small
or vanishing values. To this aim, $B_0$ is assigned dimension one in the
infra-red momenta of the external Goldstone bosons and accordingly quark
masses are also assigned dimension one. Due to this rule, at each order
of the perturbative expansion, generalized $\chi PT$ contains more terms
than standard $\chi PT$. For instance, the pion mass formula becomes, at
leading order, $m_{\pi}^2=2\hat mB_0+4\hat m^2A_0$, where the constant 
$A_0$ is expressible in terms of two-point functions of scalar and
pseudoscalar quark densities. In the standard $\chi PT$ case,
this term is relegated to the next-to-leading order.
\par
At the tree level of the chiral effective lagrangian, the $\pi\pi$ scattering
amplitude $A(s|t,u)$ has the expression
$A(s|t,u)=(s-2\hat mB_0)/F^2$,
which displays explicit dependence on the quark condensate parameter.
It is useful to introduce two parameters, $\alpha$ and $\beta$, that allow
one to express the amplitude $A$ in terms of the physical constants 
$F_{\pi}$ and $m_{\pi}$ \cite{fss,kmsf}:
\be \lb{e6}
A(s|t,u)=\frac{\beta}{F_{\pi}^2}(s-\frac{4}{3}m_{\pi}^2)+
\alpha\frac{m_{\pi}^2}{3F_{\pi}^2}.
\ee
At leading order, $F_{\pi}=F$ and $\beta=1$. The GOR
parameter (\rf{e3}) is then related to the parameter $\alpha$ by the
relation: $x_{GOR}=(4-\alpha)/3$.
For $\alpha=1$, one recovers the standard $\chi PT$ case, where $x_{GOR}=1$.
When $\alpha$ increases, $x_{GOR}$ decreases, corresponding to
decreasing values of the quark condensate. For $\alpha=4$, one reaches the
extreme case of generalized $\chi PT$, where $x_{GOR}$ and the quark
condensate vanish. At higher orders, the relationship between $\alpha$
and $x_{GOR}$ becomes more complicated, involving the parameter $\beta$
and other low energy constants, but the main qualitative feature found above
remains: values of $\alpha$ close to 1 correspond to the case of standard
$\chi PT$, while large values of $\alpha$, above 1.5-2, say, cover the
complementary domain of generalized $\chi PT$.
\par
At two-loop order, the $\pi\pi$ scattering amplitude is described by six
parameters, $\alpha$, $\beta$, $\lambda_1$, $\lambda_2$, $\lambda_3$ and
$\lambda_4$. The five parameters other than $\alpha$ are weakly dependent
on the quark condensate and the values of the $\lambda$'s are fixed by a 
detailed analysis of sum rules using available data on $\pi\pi$ scattering 
at medium energies. The parameter $\beta$ is essentially sensitive to the 
deviation of $F_{\pi}$ from $F$ and remains in general close to 1.
One ends up, in generalized $\chi PT$, with a complete determination of the
threshold parameters of the $\pi\pi$ scattering amplitude in terms of the 
possible numerical values of the parameter $\alpha$ together with a best 
value of the other parameters.
The expressions of the scattering lengths to two-loop order, as well as
the values of the parameters $\lambda$, can be found in Ref. \cite{kmsf}.
\par
The evaluation of the corrections contained in the decay width formula
(\rf{e2}) can be done in much the same way as in the standard $\chi PT$
case, with the difference that whenever scattering lengths or the quark
condensate parameter appear, these should be expressed through the
generalized $\chi PT$ formulas or parametrizations. We have followed the
same method of approach as in Refs. \cite{js,s} and skip here the
details of the calculations. We only note that we made the approximation of
taking the same values for the
electromagnetic low energy constants $k_i$ in generalized and standard
$\chi PT$ \cite{u,mms,ku}. Actually, the effective lagrangian with
electromagnetism is not yet available in generalized $\chi PT$, where the
number of such constants should be larger. However, the contributions of the
constants $k_i$ not being dominant in the total decay width correction, such
an approximation seems to be justified. Furthermore, a 100\% uncertainty
has been assigned to their global contributions in the various types of
correction, their values being taken from Ref. \cite{bu}. Also, weakly
dependent terms on the parameter $\alpha$ have been numerically taken
the same as in the standard case. Numerical estimates have been done with
the following inputs: $F_{\pi}=92.4$ MeV, $m_{\pi^+}=139.57$ MeV,
$m_{\pi^0}=134.97$ MeV.
\par
We have chosen to analyze the total correction in terms of the
scattering lengths, which are the physical quantities of interest, rather
than of the intermediate parameters $\alpha$ and $\beta$. Since the pionium
lifetime concerns at leading order the combination $(a_0^0-a_0^2)$ of the
scattering lengths, it is natural to promote this quantity as the main
variable of the problem. As a second variable, we choose the $P$-wave
scattering length, which is most sensitive to the parameter $\beta$ and
almost insensitive to $\alpha$. One then expresses the total
correction in terms of $(a_0^0-a_0^2)$, $a_1^1$ and the
$\lambda$'s. (It is sufficient for the correction term to use the one-loop
expressions of the scattering lengths, from which $\lambda_3$ and
$\lambda_4$ are absent.) 
The experimental value of $a_1^1m_{\pi}^2$ is $0.038\pm 0.002$
\cite{n}. One can then study the dependence of the decay width (or of the
lifetime) upon the combination $(a_0^0-a_0^2)$ of the scattering lengths for
fixed values of the above parameters. 
\par
The previous analysis can be repeated with various sets of inputs for
$a_1^1$ and for the $\lambda$'s within the intervals of their possible
values. Generally, for a given $(a_0^0-a_0^2)$, the lifetime $\tau$ varies
little for such changes of the inputs. The variations of the $\lambda$'s 
induce at most a variation of 0.8\% on $\tau$, while the variations of 
$a_1^1$ induce at most a variation of 1.1\%. Assuming these variations 
as being uncorrelated, one may consider that 2\% is an upper bound for the 
possible variations of $\tau$. Considering the latter as un uncertainty and
adding it to the 2\% uncertainty obtained in 
the course of calculation of $\Delta \tau/\tau_0$, one obtains a total 
uncertainty of 4\% around the value of $\tau$ calculated with the central 
values of $a_1^1$ and of the $\lambda$'s.
\par
The above analyzes are graphically summarized in Fig. \rf{f1}.
The full line represents the lifetime $\tau$ as a
function of the the combination $(a_0^0-a_0^2)$, corresponding to the
central values of $a_1^1$ and of the $\lambda$'s. The band
around it corresponds to the estimated 4\% total uncertainty.
As already mentioned, the parameter $\beta$ is almost insensitive to the
variations of $\alpha$; for instance, on the central line, it varies
between 1.10 and 1.11. For a given experimental value of the lifetime, with
a possible uncertainty, one can deduce from Fig. \rf{f1} the corresponding
value of the combination $(a_0^0-a_0^2)$ with the related uncertainty.
\par
\bfg
\setlength{\unitlength}{0.240900pt}
\ifx\plotpoint\undefined\newsavebox{\plotpoint}\fi
\sbox{\plotpoint}{\rule[-0.200pt]{0.400pt}{0.400pt}}%
\begin{picture}(825,810)(0,0)
\font\gnuplot=cmr10 at 10pt
\gnuplot
\sbox{\plotpoint}{\rule[-0.200pt]{0.400pt}{0.400pt}}%
\put(181.0,163.0){\rule[-0.200pt]{4.818pt}{0.400pt}}
\put(161,163){\makebox(0,0)[r]{1}}
\put(785.0,163.0){\rule[-0.200pt]{4.818pt}{0.400pt}}
\put(181.0,264.0){\rule[-0.200pt]{4.818pt}{0.400pt}}
\put(161,264){\makebox(0,0)[r]{1.5}}
\put(785.0,264.0){\rule[-0.200pt]{4.818pt}{0.400pt}}
\put(181.0,365.0){\rule[-0.200pt]{4.818pt}{0.400pt}}
\put(161,365){\makebox(0,0)[r]{2}}
\put(785.0,365.0){\rule[-0.200pt]{4.818pt}{0.400pt}}
\put(181.0,466.0){\rule[-0.200pt]{4.818pt}{0.400pt}}
\put(161,466){\makebox(0,0)[r]{2.5}}
\put(785.0,466.0){\rule[-0.200pt]{4.818pt}{0.400pt}}
\put(181.0,567.0){\rule[-0.200pt]{4.818pt}{0.400pt}}
\put(161,567){\makebox(0,0)[r]{3}}
\put(785.0,567.0){\rule[-0.200pt]{4.818pt}{0.400pt}}
\put(181.0,668.0){\rule[-0.200pt]{4.818pt}{0.400pt}}
\put(161,668){\makebox(0,0)[r]{3.5}}
\put(785.0,668.0){\rule[-0.200pt]{4.818pt}{0.400pt}}
\put(181.0,769.0){\rule[-0.200pt]{4.818pt}{0.400pt}}
\put(161,769){\makebox(0,0)[r]{4}}
\put(785.0,769.0){\rule[-0.200pt]{4.818pt}{0.400pt}}
\put(181.0,163.0){\rule[-0.200pt]{0.400pt}{4.818pt}}
\put(181,122){\makebox(0,0){0.25}}
\put(181.0,749.0){\rule[-0.200pt]{0.400pt}{4.818pt}}
\put(285.0,163.0){\rule[-0.200pt]{0.400pt}{4.818pt}}
\put(285,122){\makebox(0,0){0.27}}
\put(285.0,749.0){\rule[-0.200pt]{0.400pt}{4.818pt}}
\put(389.0,163.0){\rule[-0.200pt]{0.400pt}{4.818pt}}
\put(389,122){\makebox(0,0){0.29}}
\put(389.0,749.0){\rule[-0.200pt]{0.400pt}{4.818pt}}
\put(493.0,163.0){\rule[-0.200pt]{0.400pt}{4.818pt}}
\put(493,122){\makebox(0,0){0.31}}
\put(493.0,749.0){\rule[-0.200pt]{0.400pt}{4.818pt}}
\put(597.0,163.0){\rule[-0.200pt]{0.400pt}{4.818pt}}
\put(597,122){\makebox(0,0){0.33}}
\put(597.0,749.0){\rule[-0.200pt]{0.400pt}{4.818pt}}
\put(701.0,163.0){\rule[-0.200pt]{0.400pt}{4.818pt}}
\put(701,122){\makebox(0,0){0.35}}
\put(701.0,749.0){\rule[-0.200pt]{0.400pt}{4.818pt}}
\put(805.0,163.0){\rule[-0.200pt]{0.400pt}{4.818pt}}
\put(805,122){\makebox(0,0){0.37}}
\put(805.0,749.0){\rule[-0.200pt]{0.400pt}{4.818pt}}
\put(181.0,163.0){\rule[-0.200pt]{150.322pt}{0.400pt}}
\put(805.0,163.0){\rule[-0.200pt]{0.400pt}{145.985pt}}
\put(181.0,769.0){\rule[-0.200pt]{150.322pt}{0.400pt}}
\put(246,666){\makebox(0,0){$\tau$}}
\put(493,61){\makebox(0,0){$a_0^0-a_0^2$}}
\put(181.0,163.0){\rule[-0.200pt]{0.400pt}{145.985pt}}
\multiput(181.00,596.93)(0.599,-0.477){7}{\rule{0.580pt}{0.115pt}}
\multiput(181.00,597.17)(4.796,-5.000){2}{\rule{0.290pt}{0.400pt}}
\multiput(187.00,591.93)(0.581,-0.482){9}{\rule{0.567pt}{0.116pt}}
\multiput(187.00,592.17)(5.824,-6.000){2}{\rule{0.283pt}{0.400pt}}
\multiput(194.00,585.93)(0.599,-0.477){7}{\rule{0.580pt}{0.115pt}}
\multiput(194.00,586.17)(4.796,-5.000){2}{\rule{0.290pt}{0.400pt}}
\multiput(200.00,580.93)(0.491,-0.482){9}{\rule{0.500pt}{0.116pt}}
\multiput(200.00,581.17)(4.962,-6.000){2}{\rule{0.250pt}{0.400pt}}
\multiput(206.00,574.93)(0.710,-0.477){7}{\rule{0.660pt}{0.115pt}}
\multiput(206.00,575.17)(5.630,-5.000){2}{\rule{0.330pt}{0.400pt}}
\multiput(213.00,569.93)(0.599,-0.477){7}{\rule{0.580pt}{0.115pt}}
\multiput(213.00,570.17)(4.796,-5.000){2}{\rule{0.290pt}{0.400pt}}
\multiput(219.00,564.93)(0.491,-0.482){9}{\rule{0.500pt}{0.116pt}}
\multiput(219.00,565.17)(4.962,-6.000){2}{\rule{0.250pt}{0.400pt}}
\multiput(225.00,558.93)(0.599,-0.477){7}{\rule{0.580pt}{0.115pt}}
\multiput(225.00,559.17)(4.796,-5.000){2}{\rule{0.290pt}{0.400pt}}
\multiput(231.00,553.93)(0.710,-0.477){7}{\rule{0.660pt}{0.115pt}}
\multiput(231.00,554.17)(5.630,-5.000){2}{\rule{0.330pt}{0.400pt}}
\multiput(238.00,548.93)(0.599,-0.477){7}{\rule{0.580pt}{0.115pt}}
\multiput(238.00,549.17)(4.796,-5.000){2}{\rule{0.290pt}{0.400pt}}
\multiput(244.00,543.93)(0.599,-0.477){7}{\rule{0.580pt}{0.115pt}}
\multiput(244.00,544.17)(4.796,-5.000){2}{\rule{0.290pt}{0.400pt}}
\multiput(250.00,538.93)(0.710,-0.477){7}{\rule{0.660pt}{0.115pt}}
\multiput(250.00,539.17)(5.630,-5.000){2}{\rule{0.330pt}{0.400pt}}
\multiput(257.00,533.93)(0.599,-0.477){7}{\rule{0.580pt}{0.115pt}}
\multiput(257.00,534.17)(4.796,-5.000){2}{\rule{0.290pt}{0.400pt}}
\multiput(263.00,528.94)(0.774,-0.468){5}{\rule{0.700pt}{0.113pt}}
\multiput(263.00,529.17)(4.547,-4.000){2}{\rule{0.350pt}{0.400pt}}
\multiput(269.00,524.93)(0.710,-0.477){7}{\rule{0.660pt}{0.115pt}}
\multiput(269.00,525.17)(5.630,-5.000){2}{\rule{0.330pt}{0.400pt}}
\multiput(276.00,519.93)(0.599,-0.477){7}{\rule{0.580pt}{0.115pt}}
\multiput(276.00,520.17)(4.796,-5.000){2}{\rule{0.290pt}{0.400pt}}
\multiput(282.00,514.94)(0.774,-0.468){5}{\rule{0.700pt}{0.113pt}}
\multiput(282.00,515.17)(4.547,-4.000){2}{\rule{0.350pt}{0.400pt}}
\multiput(288.00,510.93)(0.599,-0.477){7}{\rule{0.580pt}{0.115pt}}
\multiput(288.00,511.17)(4.796,-5.000){2}{\rule{0.290pt}{0.400pt}}
\multiput(294.00,505.94)(0.920,-0.468){5}{\rule{0.800pt}{0.113pt}}
\multiput(294.00,506.17)(5.340,-4.000){2}{\rule{0.400pt}{0.400pt}}
\multiput(301.00,501.93)(0.599,-0.477){7}{\rule{0.580pt}{0.115pt}}
\multiput(301.00,502.17)(4.796,-5.000){2}{\rule{0.290pt}{0.400pt}}
\multiput(307.00,496.94)(0.774,-0.468){5}{\rule{0.700pt}{0.113pt}}
\multiput(307.00,497.17)(4.547,-4.000){2}{\rule{0.350pt}{0.400pt}}
\multiput(313.00,492.94)(0.920,-0.468){5}{\rule{0.800pt}{0.113pt}}
\multiput(313.00,493.17)(5.340,-4.000){2}{\rule{0.400pt}{0.400pt}}
\multiput(320.00,488.93)(0.599,-0.477){7}{\rule{0.580pt}{0.115pt}}
\multiput(320.00,489.17)(4.796,-5.000){2}{\rule{0.290pt}{0.400pt}}
\multiput(326.00,483.94)(0.774,-0.468){5}{\rule{0.700pt}{0.113pt}}
\multiput(326.00,484.17)(4.547,-4.000){2}{\rule{0.350pt}{0.400pt}}
\multiput(332.00,479.94)(0.920,-0.468){5}{\rule{0.800pt}{0.113pt}}
\multiput(332.00,480.17)(5.340,-4.000){2}{\rule{0.400pt}{0.400pt}}
\multiput(339.00,475.94)(0.774,-0.468){5}{\rule{0.700pt}{0.113pt}}
\multiput(339.00,476.17)(4.547,-4.000){2}{\rule{0.350pt}{0.400pt}}
\multiput(345.00,471.94)(0.774,-0.468){5}{\rule{0.700pt}{0.113pt}}
\multiput(345.00,472.17)(4.547,-4.000){2}{\rule{0.350pt}{0.400pt}}
\multiput(351.00,467.94)(0.920,-0.468){5}{\rule{0.800pt}{0.113pt}}
\multiput(351.00,468.17)(5.340,-4.000){2}{\rule{0.400pt}{0.400pt}}
\multiput(358.00,463.94)(0.774,-0.468){5}{\rule{0.700pt}{0.113pt}}
\multiput(358.00,464.17)(4.547,-4.000){2}{\rule{0.350pt}{0.400pt}}
\multiput(364.00,459.94)(0.774,-0.468){5}{\rule{0.700pt}{0.113pt}}
\multiput(364.00,460.17)(4.547,-4.000){2}{\rule{0.350pt}{0.400pt}}
\multiput(370.00,455.94)(0.774,-0.468){5}{\rule{0.700pt}{0.113pt}}
\multiput(370.00,456.17)(4.547,-4.000){2}{\rule{0.350pt}{0.400pt}}
\multiput(376.00,451.94)(0.920,-0.468){5}{\rule{0.800pt}{0.113pt}}
\multiput(376.00,452.17)(5.340,-4.000){2}{\rule{0.400pt}{0.400pt}}
\multiput(383.00,447.94)(0.774,-0.468){5}{\rule{0.700pt}{0.113pt}}
\multiput(383.00,448.17)(4.547,-4.000){2}{\rule{0.350pt}{0.400pt}}
\multiput(389.00,443.95)(1.132,-0.447){3}{\rule{0.900pt}{0.108pt}}
\multiput(389.00,444.17)(4.132,-3.000){2}{\rule{0.450pt}{0.400pt}}
\multiput(395.00,440.94)(0.920,-0.468){5}{\rule{0.800pt}{0.113pt}}
\multiput(395.00,441.17)(5.340,-4.000){2}{\rule{0.400pt}{0.400pt}}
\multiput(402.00,436.94)(0.774,-0.468){5}{\rule{0.700pt}{0.113pt}}
\multiput(402.00,437.17)(4.547,-4.000){2}{\rule{0.350pt}{0.400pt}}
\multiput(408.00,432.95)(1.132,-0.447){3}{\rule{0.900pt}{0.108pt}}
\multiput(408.00,433.17)(4.132,-3.000){2}{\rule{0.450pt}{0.400pt}}
\multiput(414.00,429.94)(0.920,-0.468){5}{\rule{0.800pt}{0.113pt}}
\multiput(414.00,430.17)(5.340,-4.000){2}{\rule{0.400pt}{0.400pt}}
\multiput(421.00,425.95)(1.132,-0.447){3}{\rule{0.900pt}{0.108pt}}
\multiput(421.00,426.17)(4.132,-3.000){2}{\rule{0.450pt}{0.400pt}}
\multiput(427.00,422.94)(0.774,-0.468){5}{\rule{0.700pt}{0.113pt}}
\multiput(427.00,423.17)(4.547,-4.000){2}{\rule{0.350pt}{0.400pt}}
\multiput(433.00,418.95)(1.132,-0.447){3}{\rule{0.900pt}{0.108pt}}
\multiput(433.00,419.17)(4.132,-3.000){2}{\rule{0.450pt}{0.400pt}}
\multiput(439.00,415.94)(0.920,-0.468){5}{\rule{0.800pt}{0.113pt}}
\multiput(439.00,416.17)(5.340,-4.000){2}{\rule{0.400pt}{0.400pt}}
\multiput(446.00,411.95)(1.132,-0.447){3}{\rule{0.900pt}{0.108pt}}
\multiput(446.00,412.17)(4.132,-3.000){2}{\rule{0.450pt}{0.400pt}}
\multiput(452.00,408.94)(0.774,-0.468){5}{\rule{0.700pt}{0.113pt}}
\multiput(452.00,409.17)(4.547,-4.000){2}{\rule{0.350pt}{0.400pt}}
\multiput(458.00,404.95)(1.355,-0.447){3}{\rule{1.033pt}{0.108pt}}
\multiput(458.00,405.17)(4.855,-3.000){2}{\rule{0.517pt}{0.400pt}}
\multiput(465.00,401.95)(1.132,-0.447){3}{\rule{0.900pt}{0.108pt}}
\multiput(465.00,402.17)(4.132,-3.000){2}{\rule{0.450pt}{0.400pt}}
\multiput(471.00,398.95)(1.132,-0.447){3}{\rule{0.900pt}{0.108pt}}
\multiput(471.00,399.17)(4.132,-3.000){2}{\rule{0.450pt}{0.400pt}}
\multiput(477.00,395.94)(0.920,-0.468){5}{\rule{0.800pt}{0.113pt}}
\multiput(477.00,396.17)(5.340,-4.000){2}{\rule{0.400pt}{0.400pt}}
\multiput(484.00,391.95)(1.132,-0.447){3}{\rule{0.900pt}{0.108pt}}
\multiput(484.00,392.17)(4.132,-3.000){2}{\rule{0.450pt}{0.400pt}}
\multiput(490.00,388.95)(1.132,-0.447){3}{\rule{0.900pt}{0.108pt}}
\multiput(490.00,389.17)(4.132,-3.000){2}{\rule{0.450pt}{0.400pt}}
\multiput(496.00,385.95)(1.355,-0.447){3}{\rule{1.033pt}{0.108pt}}
\multiput(496.00,386.17)(4.855,-3.000){2}{\rule{0.517pt}{0.400pt}}
\multiput(503.00,382.95)(1.132,-0.447){3}{\rule{0.900pt}{0.108pt}}
\multiput(503.00,383.17)(4.132,-3.000){2}{\rule{0.450pt}{0.400pt}}
\multiput(509.00,379.95)(1.132,-0.447){3}{\rule{0.900pt}{0.108pt}}
\multiput(509.00,380.17)(4.132,-3.000){2}{\rule{0.450pt}{0.400pt}}
\multiput(515.00,376.95)(1.132,-0.447){3}{\rule{0.900pt}{0.108pt}}
\multiput(515.00,377.17)(4.132,-3.000){2}{\rule{0.450pt}{0.400pt}}
\multiput(521.00,373.95)(1.355,-0.447){3}{\rule{1.033pt}{0.108pt}}
\multiput(521.00,374.17)(4.855,-3.000){2}{\rule{0.517pt}{0.400pt}}
\multiput(528.00,370.95)(1.132,-0.447){3}{\rule{0.900pt}{0.108pt}}
\multiput(528.00,371.17)(4.132,-3.000){2}{\rule{0.450pt}{0.400pt}}
\multiput(534.00,367.95)(1.132,-0.447){3}{\rule{0.900pt}{0.108pt}}
\multiput(534.00,368.17)(4.132,-3.000){2}{\rule{0.450pt}{0.400pt}}
\multiput(540.00,364.95)(1.355,-0.447){3}{\rule{1.033pt}{0.108pt}}
\multiput(540.00,365.17)(4.855,-3.000){2}{\rule{0.517pt}{0.400pt}}
\multiput(547.00,361.95)(1.132,-0.447){3}{\rule{0.900pt}{0.108pt}}
\multiput(547.00,362.17)(4.132,-3.000){2}{\rule{0.450pt}{0.400pt}}
\put(553,358.17){\rule{1.300pt}{0.400pt}}
\multiput(553.00,359.17)(3.302,-2.000){2}{\rule{0.650pt}{0.400pt}}
\multiput(559.00,356.95)(1.355,-0.447){3}{\rule{1.033pt}{0.108pt}}
\multiput(559.00,357.17)(4.855,-3.000){2}{\rule{0.517pt}{0.400pt}}
\multiput(566.00,353.95)(1.132,-0.447){3}{\rule{0.900pt}{0.108pt}}
\multiput(566.00,354.17)(4.132,-3.000){2}{\rule{0.450pt}{0.400pt}}
\multiput(572.00,350.95)(1.132,-0.447){3}{\rule{0.900pt}{0.108pt}}
\multiput(572.00,351.17)(4.132,-3.000){2}{\rule{0.450pt}{0.400pt}}
\put(578,347.17){\rule{1.500pt}{0.400pt}}
\multiput(578.00,348.17)(3.887,-2.000){2}{\rule{0.750pt}{0.400pt}}
\multiput(585.00,345.95)(1.132,-0.447){3}{\rule{0.900pt}{0.108pt}}
\multiput(585.00,346.17)(4.132,-3.000){2}{\rule{0.450pt}{0.400pt}}
\multiput(591.00,342.95)(1.132,-0.447){3}{\rule{0.900pt}{0.108pt}}
\multiput(591.00,343.17)(4.132,-3.000){2}{\rule{0.450pt}{0.400pt}}
\put(597,339.17){\rule{1.300pt}{0.400pt}}
\multiput(597.00,340.17)(3.302,-2.000){2}{\rule{0.650pt}{0.400pt}}
\multiput(603.00,337.95)(1.355,-0.447){3}{\rule{1.033pt}{0.108pt}}
\multiput(603.00,338.17)(4.855,-3.000){2}{\rule{0.517pt}{0.400pt}}
\multiput(610.00,334.95)(1.132,-0.447){3}{\rule{0.900pt}{0.108pt}}
\multiput(610.00,335.17)(4.132,-3.000){2}{\rule{0.450pt}{0.400pt}}
\put(616,331.17){\rule{1.300pt}{0.400pt}}
\multiput(616.00,332.17)(3.302,-2.000){2}{\rule{0.650pt}{0.400pt}}
\multiput(622.00,329.95)(1.355,-0.447){3}{\rule{1.033pt}{0.108pt}}
\multiput(622.00,330.17)(4.855,-3.000){2}{\rule{0.517pt}{0.400pt}}
\put(629,326.17){\rule{1.300pt}{0.400pt}}
\multiput(629.00,327.17)(3.302,-2.000){2}{\rule{0.650pt}{0.400pt}}
\multiput(635.00,324.95)(1.132,-0.447){3}{\rule{0.900pt}{0.108pt}}
\multiput(635.00,325.17)(4.132,-3.000){2}{\rule{0.450pt}{0.400pt}}
\put(641,321.17){\rule{1.500pt}{0.400pt}}
\multiput(641.00,322.17)(3.887,-2.000){2}{\rule{0.750pt}{0.400pt}}
\put(648,319.17){\rule{1.300pt}{0.400pt}}
\multiput(648.00,320.17)(3.302,-2.000){2}{\rule{0.650pt}{0.400pt}}
\multiput(654.00,317.95)(1.132,-0.447){3}{\rule{0.900pt}{0.108pt}}
\multiput(654.00,318.17)(4.132,-3.000){2}{\rule{0.450pt}{0.400pt}}
\put(660,314.17){\rule{1.300pt}{0.400pt}}
\multiput(660.00,315.17)(3.302,-2.000){2}{\rule{0.650pt}{0.400pt}}
\put(666,312.17){\rule{1.500pt}{0.400pt}}
\multiput(666.00,313.17)(3.887,-2.000){2}{\rule{0.750pt}{0.400pt}}
\multiput(673.00,310.95)(1.132,-0.447){3}{\rule{0.900pt}{0.108pt}}
\multiput(673.00,311.17)(4.132,-3.000){2}{\rule{0.450pt}{0.400pt}}
\put(679,307.17){\rule{1.300pt}{0.400pt}}
\multiput(679.00,308.17)(3.302,-2.000){2}{\rule{0.650pt}{0.400pt}}
\put(685,305.17){\rule{1.500pt}{0.400pt}}
\multiput(685.00,306.17)(3.887,-2.000){2}{\rule{0.750pt}{0.400pt}}
\multiput(692.00,303.95)(1.132,-0.447){3}{\rule{0.900pt}{0.108pt}}
\multiput(692.00,304.17)(4.132,-3.000){2}{\rule{0.450pt}{0.400pt}}
\put(698,300.17){\rule{1.300pt}{0.400pt}}
\multiput(698.00,301.17)(3.302,-2.000){2}{\rule{0.650pt}{0.400pt}}
\put(704,298.17){\rule{1.500pt}{0.400pt}}
\multiput(704.00,299.17)(3.887,-2.000){2}{\rule{0.750pt}{0.400pt}}
\put(711,296.17){\rule{1.300pt}{0.400pt}}
\multiput(711.00,297.17)(3.302,-2.000){2}{\rule{0.650pt}{0.400pt}}
\put(717,294.17){\rule{1.300pt}{0.400pt}}
\multiput(717.00,295.17)(3.302,-2.000){2}{\rule{0.650pt}{0.400pt}}
\multiput(723.00,292.95)(1.355,-0.447){3}{\rule{1.033pt}{0.108pt}}
\multiput(723.00,293.17)(4.855,-3.000){2}{\rule{0.517pt}{0.400pt}}
\put(730,289.17){\rule{1.300pt}{0.400pt}}
\multiput(730.00,290.17)(3.302,-2.000){2}{\rule{0.650pt}{0.400pt}}
\put(736,287.17){\rule{1.300pt}{0.400pt}}
\multiput(736.00,288.17)(3.302,-2.000){2}{\rule{0.650pt}{0.400pt}}
\put(742,285.17){\rule{1.300pt}{0.400pt}}
\multiput(742.00,286.17)(3.302,-2.000){2}{\rule{0.650pt}{0.400pt}}
\put(748,283.17){\rule{1.500pt}{0.400pt}}
\multiput(748.00,284.17)(3.887,-2.000){2}{\rule{0.750pt}{0.400pt}}
\put(755,281.17){\rule{1.300pt}{0.400pt}}
\multiput(755.00,282.17)(3.302,-2.000){2}{\rule{0.650pt}{0.400pt}}
\put(761,279.17){\rule{1.300pt}{0.400pt}}
\multiput(761.00,280.17)(3.302,-2.000){2}{\rule{0.650pt}{0.400pt}}
\put(767,277.17){\rule{1.500pt}{0.400pt}}
\multiput(767.00,278.17)(3.887,-2.000){2}{\rule{0.750pt}{0.400pt}}
\put(774,275.17){\rule{1.300pt}{0.400pt}}
\multiput(774.00,276.17)(3.302,-2.000){2}{\rule{0.650pt}{0.400pt}}
\put(780,273.17){\rule{1.300pt}{0.400pt}}
\multiput(780.00,274.17)(3.302,-2.000){2}{\rule{0.650pt}{0.400pt}}
\put(786,271.17){\rule{1.500pt}{0.400pt}}
\multiput(786.00,272.17)(3.887,-2.000){2}{\rule{0.750pt}{0.400pt}}
\put(793,269.17){\rule{1.300pt}{0.400pt}}
\multiput(793.00,270.17)(3.302,-2.000){2}{\rule{0.650pt}{0.400pt}}
\put(799,267.17){\rule{1.300pt}{0.400pt}}
\multiput(799.00,268.17)(3.302,-2.000){2}{\rule{0.650pt}{0.400pt}}
\sbox{\plotpoint}{\rule[-0.500pt]{1.000pt}{1.000pt}}%
\put(181.00,626.00){\usebox{\plotpoint}}
\put(196.16,611.84){\usebox{\plotpoint}}
\put(211.70,598.11){\usebox{\plotpoint}}
\put(227.11,584.24){\usebox{\plotpoint}}
\put(242.98,570.85){\usebox{\plotpoint}}
\put(259.31,558.07){\usebox{\plotpoint}}
\put(275.63,545.27){\usebox{\plotpoint}}
\put(291.89,532.41){\usebox{\plotpoint}}
\put(308.51,520.00){\usebox{\plotpoint}}
\put(325.62,508.26){\usebox{\plotpoint}}
\put(342.68,496.55){\usebox{\plotpoint}}
\put(359.74,484.84){\usebox{\plotpoint}}
\put(377.06,473.40){\usebox{\plotpoint}}
\put(394.57,462.28){\usebox{\plotpoint}}
\put(412.14,451.24){\usebox{\plotpoint}}
\put(430.07,440.95){\usebox{\plotpoint}}
\put(448.20,430.90){\usebox{\plotpoint}}
\put(466.40,421.07){\usebox{\plotpoint}}
\put(484.41,410.80){\usebox{\plotpoint}}
\put(503.15,401.90){\usebox{\plotpoint}}
\put(521.28,391.88){\usebox{\plotpoint}}
\put(540.03,382.99){\usebox{\plotpoint}}
\put(558.78,374.11){\usebox{\plotpoint}}
\put(577.53,365.23){\usebox{\plotpoint}}
\put(596.59,357.21){\usebox{\plotpoint}}
\put(615.64,349.18){\usebox{\plotpoint}}
\put(634.69,341.15){\usebox{\plotpoint}}
\put(653.74,333.13){\usebox{\plotpoint}}
\put(673.15,325.95){\usebox{\plotpoint}}
\put(692.54,318.73){\usebox{\plotpoint}}
\put(711.99,311.67){\usebox{\plotpoint}}
\put(731.41,304.53){\usebox{\plotpoint}}
\put(750.77,297.21){\usebox{\plotpoint}}
\put(770.57,290.98){\usebox{\plotpoint}}
\put(790.36,284.75){\usebox{\plotpoint}}
\put(805,280){\usebox{\plotpoint}}
\put(181.00,571.00){\usebox{\plotpoint}}
\put(196.82,557.65){\usebox{\plotpoint}}
\put(213.15,544.87){\usebox{\plotpoint}}
\put(229.10,531.58){\usebox{\plotpoint}}
\put(245.44,518.80){\usebox{\plotpoint}}
\put(262.19,506.68){\usebox{\plotpoint}}
\put(279.23,494.85){\usebox{\plotpoint}}
\put(296.09,482.81){\usebox{\plotpoint}}
\put(313.07,470.96){\usebox{\plotpoint}}
\put(330.63,459.92){\usebox{\plotpoint}}
\put(348.19,448.87){\usebox{\plotpoint}}
\put(366.12,438.59){\usebox{\plotpoint}}
\put(384.05,428.30){\usebox{\plotpoint}}
\put(401.99,418.01){\usebox{\plotpoint}}
\put(419.92,407.62){\usebox{\plotpoint}}
\put(438.07,397.62){\usebox{\plotpoint}}
\put(456.76,388.62){\usebox{\plotpoint}}
\put(475.06,378.97){\usebox{\plotpoint}}
\put(493.81,370.09){\usebox{\plotpoint}}
\put(512.78,361.74){\usebox{\plotpoint}}
\put(531.66,353.17){\usebox{\plotpoint}}
\put(550.62,344.79){\usebox{\plotpoint}}
\put(569.72,336.76){\usebox{\plotpoint}}
\put(588.90,329.05){\usebox{\plotpoint}}
\put(608.17,321.52){\usebox{\plotpoint}}
\put(627.35,313.71){\usebox{\plotpoint}}
\put(646.70,306.37){\usebox{\plotpoint}}
\put(666.04,298.99){\usebox{\plotpoint}}
\put(685.80,292.66){\usebox{\plotpoint}}
\put(705.31,285.63){\usebox{\plotpoint}}
\put(725.11,279.40){\usebox{\plotpoint}}
\put(744.86,273.05){\usebox{\plotpoint}}
\put(764.64,266.79){\usebox{\plotpoint}}
\put(784.43,260.52){\usebox{\plotpoint}}
\put(804.42,255.10){\usebox{\plotpoint}}
\put(805,255){\usebox{\plotpoint}}
\end{picture}
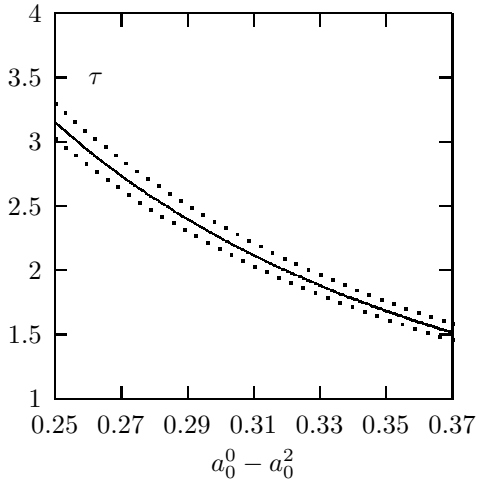
\caption{The pionium lifetime, in units of $10^{-15}$ s, as a function of 
the combination $(a_0^0-a_0^2)$ of the $S$-wave scattering lengths. The full
line corresponds to the central value $a_1^1m_{\pi}^2=0.038$ of the $P$-wave
scattering length. The band delineated by the dotted lines takes account of
the uncertainties, coming from theoretical evaluations, low energy constants 
and $a_1^1$.}
\lb{f1}
\efg
The interpretation of the experimental value of the pionium lifetime will
depend on its order of magnitude. We assume in the following discussion that,
as the DIRAC experimental project foresees, the experimental uncertainty
on the lifetime is of 10\%. For larger uncertainties, the discussion should
accordingly be modified. Three different possibilities may be considered.
First, the central value of the lifetime is close to $3\times 10^{-15}$ s,
lying above $2.9\times 10^{-15}$ s, say. Then standard $\chi PT$ is firmly
confirmed, since its predictions of $(a_0^0-a_0^2)$ lie between 0.250 and
0.258 \cite{bcegs}. Second, the central value of the lifetime lies below
$2.4\times 10^{-15}$ s. Then it is the scheme of generalized $\chi PT$ which
is confirmed, since the corresponding central values of $\alpha$ would lie
above 2, which would mean that the quark condensate is not as large as
assumed in the standard scheme. The third possibility is the most difficult
to interpret. If the central value of the lifetime lies in the interval
$2.4-2.9\times 10^{-15}$ s, then, because of the uncertainties, ambiguities
in the interpretation may arise. Let us consider as an illustrative
possibility the hypothetical result $\tau=(2.60\pm 0.26)\times 10^{-15}$ s.
Without the inclusion of the uncertainties discussed above, this would imply
$(a_0^0-a_0^2)=0.278\mp 0.015$ and $\alpha=1.68\mp 0.42$. Taking into 
account the 4\% uncertainties
represented by the band, one would have: $(a_0^0-a_0^2)=0.278\mp 0.022$
and $\alpha=1.68\mp 0.61$. Clearly, the upper bound of $\tau$ is in favor
of standard $\chi PT$, while its lower bound in favor of generalized
$\chi PT$. Such a situation would not disentangle the two schemes from
each other. The uncertainties might be reduced in this case by a more
refined analysis, taking into account the correlations between variations
of $a_1^1$ and of the $\lambda$'s.
\par
I thank J. Gasser, L. Girlanda, M. Knecht, P. Minkowski, B. Moussallam, 
L. Nemenov, A. Rusetsky and J. Stern for useful discussions.
\par

\end{document}